\documentclass[%
 aip,
 amsmath,amssymb,
 hypperef,
 reprint,%
]{revtex4-1}

\usepackage[colorlinks,linkcolor=blue,citecolor=blue]{hyperref}

\usepackage{graphicx}% Include figure files
\usepackage{dcolumn}% Align table columns on decimal point
\usepackage{bm}% bold math
\usepackage[utf8]{inputenc}
\usepackage[T1]{fontenc}
\usepackage{mathptmx}

\begin{document}

\preprint{AIP/123-QED}

%\title{Online Multi-Objective Particle Accelerator Optimization Algorithm Demonstrated at the AWAKE Electron Beam Line for Simultaneous Emittance and Orbit Control}
\title{Online Multi-Objective Particle Accelerator Optimization of the AWAKE Electron Beam Line for Simultaneous Emittance and Orbit Control}
% Force line breaks with \\

\author{Alexander~Scheinker}
\email[]{ascheink@lanl.gov}
\affiliation{Los Alamos National Laboratory, Los Alamos, New Mexico 87544, USA}
\author{Simon~Hirlaender}
\affiliation{CERN, Espl. des Particules 1, Geneva, Switzerland}
\author{Francesco~Maria~Velotti}
\affiliation{CERN, Espl. des Particules 1, Geneva, Switzerland}
\author{Spencer~Gessner} 
\affiliation{CERN, Espl. des Particules 1, Geneva, Switzerland}
\author{Giovanni~Zevi~Della~Porta} 
\affiliation{CERN, Espl. des Particules 1, Geneva, Switzerland}
\author{Verena~Kain} 
\affiliation{CERN, Espl. des Particules 1, Geneva, Switzerland}
\author{Rebecca~Ramjiawan} 
\affiliation{CERN, Espl. des Particules 1, Geneva, Switzerland}

\date{\today}% It is always \today, today,

\begin{abstract}
Multi-objective optimization is important for particle accelerators where various competing objectives must be satisfied routinely such as, for example, transverse emittance vs bunch length. We develop and demonstrate an online multi-time scale multi-objective optimization algorithm that performs real time feedback on particle accelerators. We demonstrate the ability to simultaneously minimize the emittance and maintain a reference trajectory of a beam in the electron beamline in CERN's Advanced Proton Driven Plasma Wakefield Acceleration Experiment (AWAKE).
\end{abstract}

\maketitle

\begin{figure*}[!t]
\centering
\includegraphics[width=.75\textwidth]{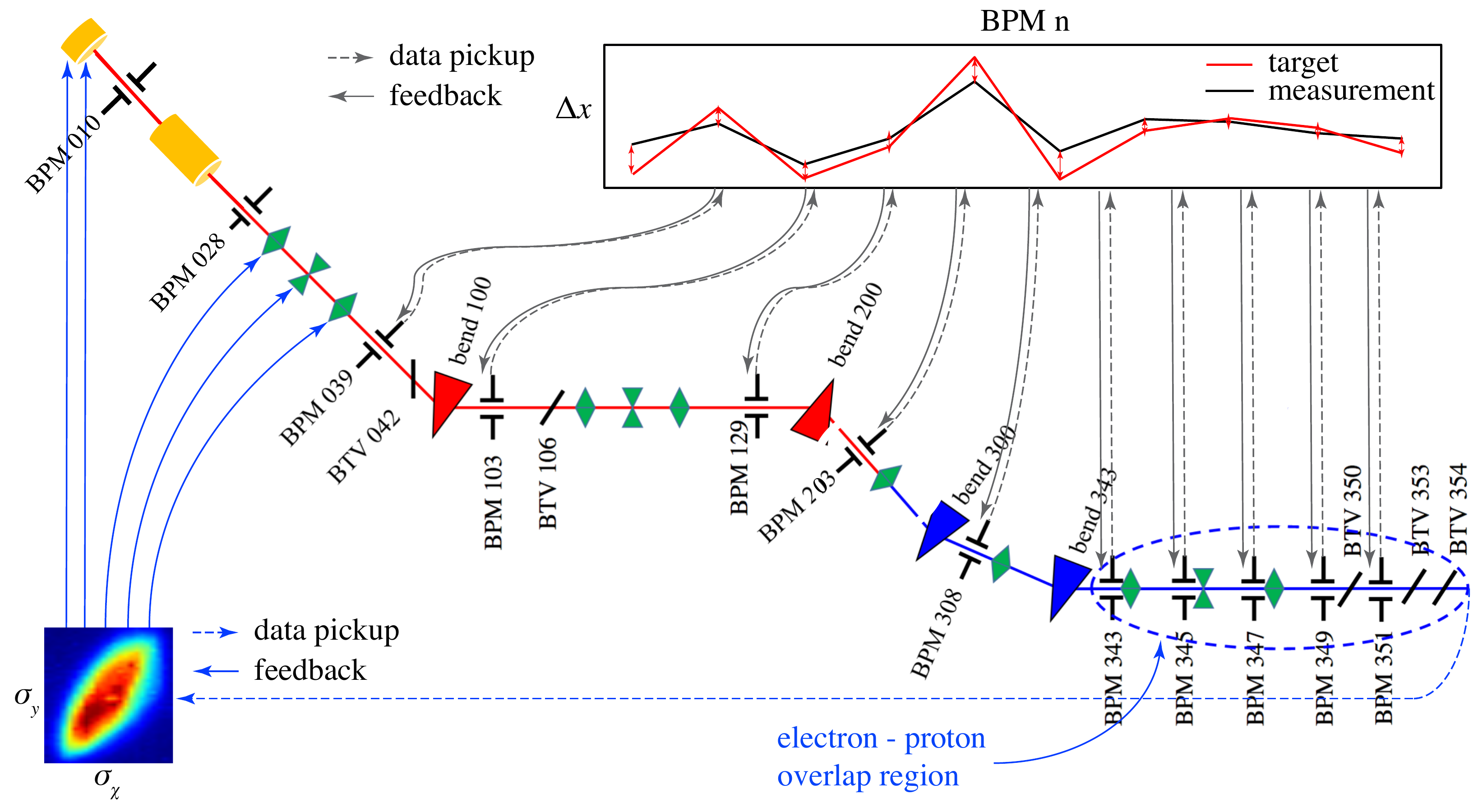}
\caption{AWAKE electron beam line and setup of two simultaneous adaptive feedback control schemes. One algorithm slowly adjusts solenoid magnet currents and quadrupole magnet settings near the injector to minimize beam size. The other feedback monitors beam position monitors and quickly adjusts steering magnets in order to maintain a design orbit despite perturbations introduced by the actions of the slower algorithm.}
\label{fig:AWAKE_e}
\end{figure*}

\begin{figure}[!t]
\centering
\includegraphics[width=.4\textwidth]{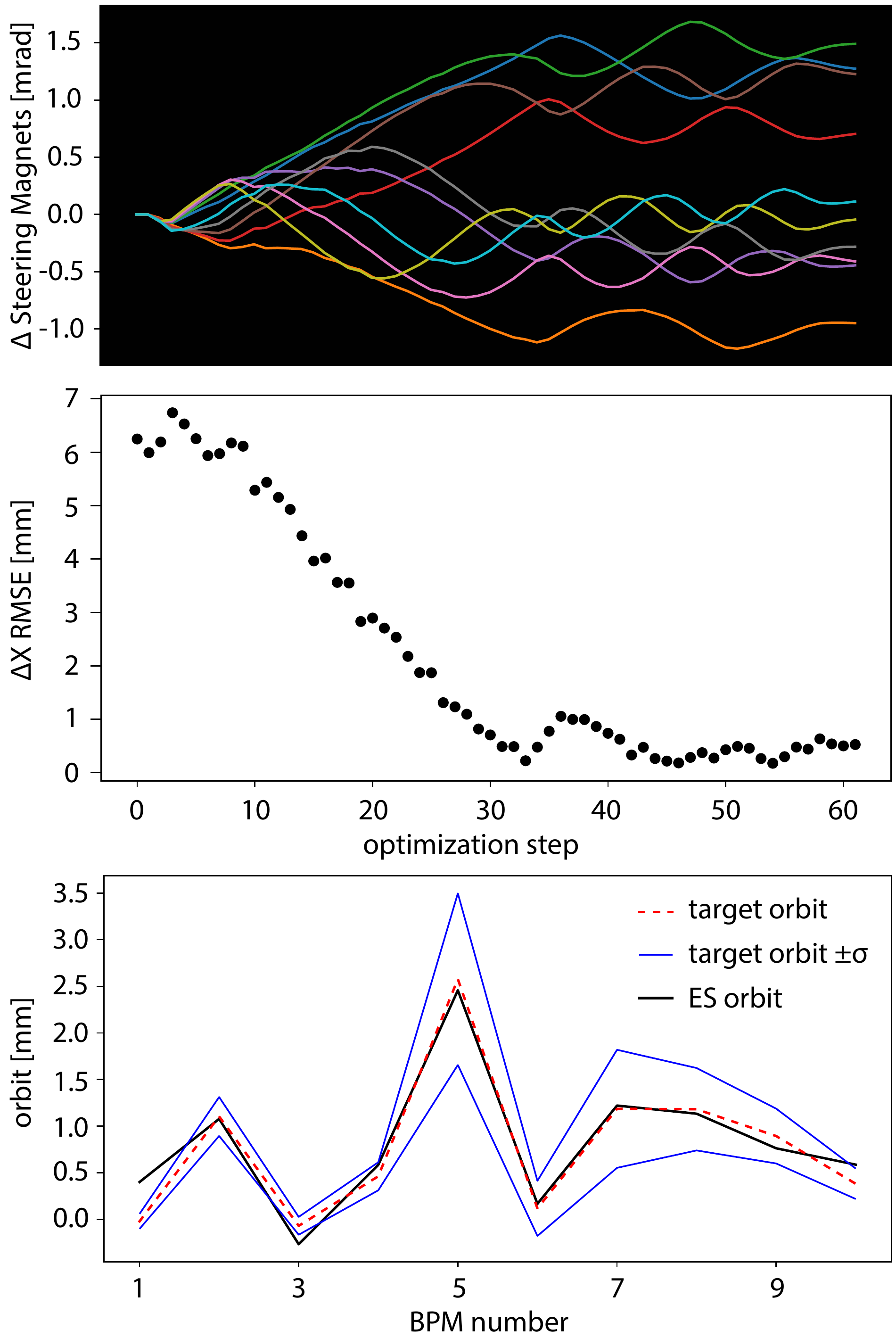}
\caption{Evolution of the 10 steering magnets is shown (top) relative to the evolution of the RMS error (middle). The bottom figure shows one snapshot of the final trajectory match relative to a target orbit based on recording and averaging 1000 BPM readings while the magnets are unchanged. The blue envelopes show $\pm \sigma$ of the orbit based on 1000 BPM readings, due to large energy jitters being introduced by the Klystron and causing trajectory deviation.}
\label{fig:traj_e}
\end{figure}

\begin{figure}[!t]
\centering
\includegraphics[width=.45\textwidth]{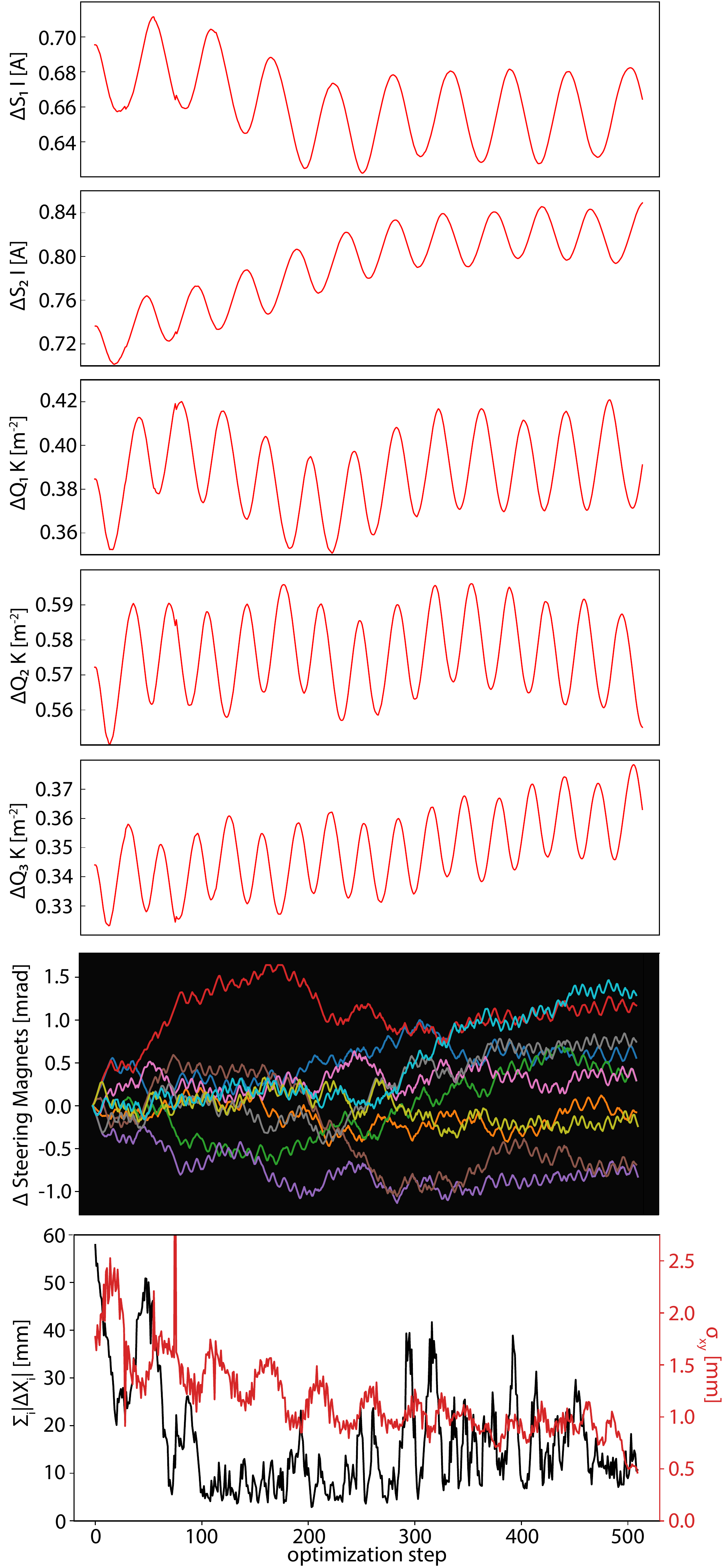}
\caption{Simultaneous evolution of the two parts of the algorithm, adjusting longitudinal and transverse beam properties is shown.}
\label{fig:emittance_params_e}
\end{figure}

\begin{figure*}[!t]
\centering
\includegraphics[width=0.85\textwidth]{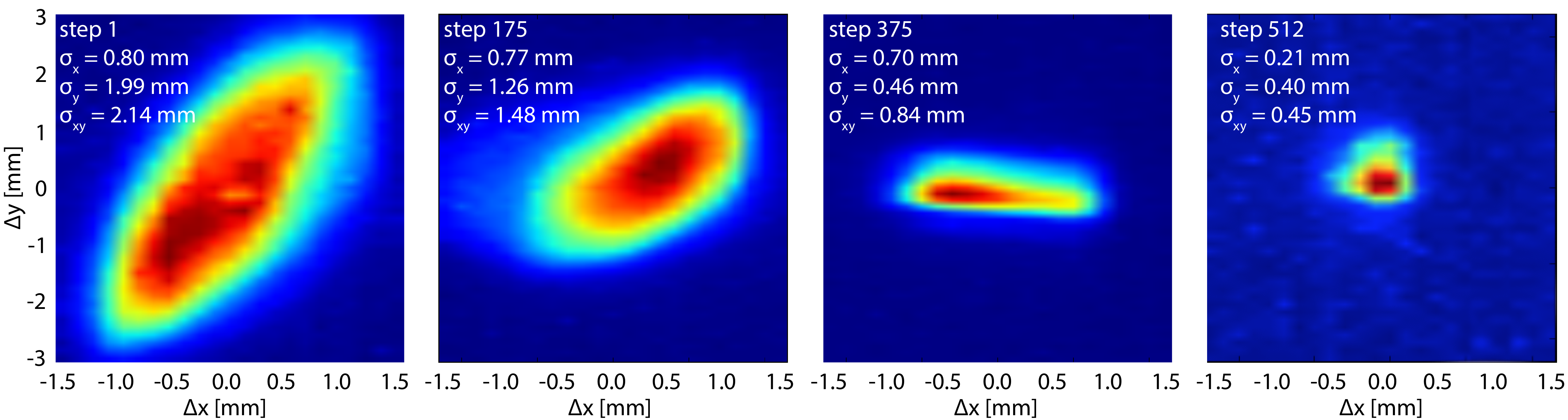}
\caption{AWAKE electron beam size at the end of the beamline, at different stages of the optimization.}
\label{fig:emittance_e}
\end{figure*}

\begin{figure}[!t]
\centering
\includegraphics[width=0.425\textwidth]{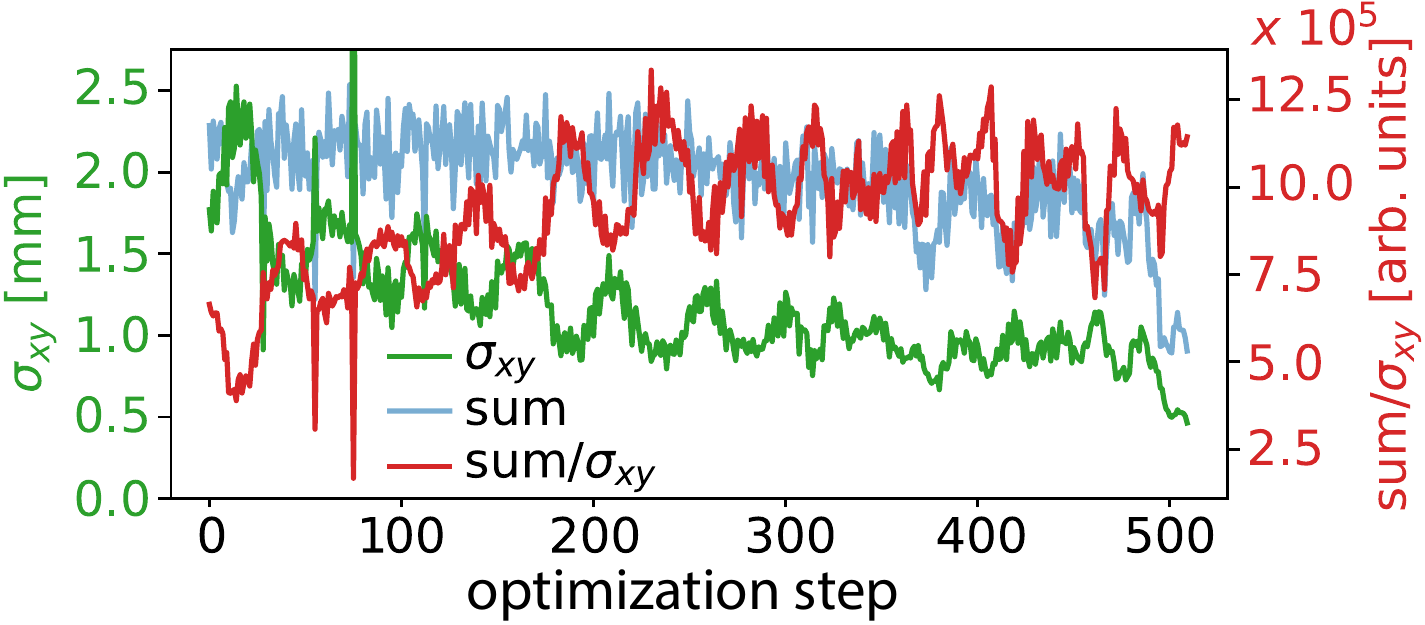}
\caption{Electron beam size and images sums used as an estimate of beam loss. Sum divided by $\sigma_{xy}$ is a measure of intensity.}
\label{fig:emit_size}
\end{figure}

High energy particle accelerators are extremely powerful scientific tools, generating intense and extremely short charged particle beams and flashes of light for a wide range of research including material science and biology. Due to their size (typically > 1 km in length) and thousands of coupled components, accelerator operation, tuning, and optimization is challenging. Accelerators utilize magnets and their associated power sources to keep charged particle beams from diverging transversely and electromagnetic resonant radio frequency (RF) cavities and their associated RF amplifiers to focus and accelerate charged particle bunches longitudinally. The control of charged particle beams is a problem that lives in a 6 dimensional phase space given by $(x,y,z,x',y',E)$ where $z = (s - ct)$ with $s$ the laboratory frame longitudinal coordinate along the axis of the accelerator, $(x,y)$ the transverse particle location off axis, $x'=dx/ds,y'=dy/ds$ measures of divergence, and $E$ particle energy. These 6 dimensions are all coupled through various electromagnetic accelerator components and through the fields of the beams themselves. 

Furthermore, both the components and the 6D distributions of the beams entering accelerators drift and change with time in unexpected ways and require continuous re-adjustment to maintain high beam quality. Because of their complexity, the performance of these machines is not easily quantified by a single number, there is usually a trade off between multiple objectives. For example, on a photocathode a trade off must be made between transverse emittance and bunch length because shorter bunches have higher charge density and therefore experience higher space charge forces that cause divergence \cite{ref_MO_opt_e}. Recently, multi-objective optimization approaches have been studied in simulation for RF cavity shape optimization to maximize shunt impedance while minimizing peak surface electric field \cite{ref_MO_opt_RF}, for electron beam dynamics simulations of the Argonne Wakefield Accelerator Facility (AWA) \cite{ref_MO_opt_AWA}, and for 3D beam tracking in electrostatic beamlines \cite{ref_MO_opt_3D}.

Such trade offs and control difficulties are present in all accelerators and require new advanced algorithms to quickly, accurately, automatically control and optimize charged particle beams. Machine learning (ML) methods have been used to develop neural network-based longitudinal phase space (LPS), $(z,E)$, diagnostics \cite{ref_LPS_ML}. An automated algorithm has also been developed and tested in hardware on the CERN Super Proton Synchrotorn (SPS) for automated septum alignment \cite{ref_SPS_opt}. Work has also begun on combining model-independent feedback algorithms with ML techniques to automatically control the LPS of charged particle beams, using the data-based ML methods  to move quickly over large distances in parameter space, and using model-independent feedback methods to zoom in on optimal settings and to maintain them despite uncertainty and time variation of beams and components \cite{ref_ES_ML_FEL}.

The need for advanced control methods exists in all accelerators due to component and beam drift. A particular class of accelerators which could greatly benefit from fast, automated tuning are free electron lasers (FEL) which are extremely flexible machines providing flashes of light many orders of magnitude brighter than other existing light sources over a wide range of photon energies and pulse lengths, and are developing advanced techniques such as self seeding \cite{ref_LCLS_FreshSlice,ref_LCLS_SS}. For example, the Linac Coherent Light Source \cite{ref_LCLS} (LCLS) FEL provides a 0.28 to 11.2 keV photon energy range with 10 - 50 fs pulse lengths, and the Swiss FEL \cite{ref_Swiss_FEL} a 1.77 to 12.4 keV photon energy range with 0.2 - 20 fs pulse lengths. Another example is plasma wakefield accelerators (PWFA), which are a class of machines that are studying methods to achieve an energy gain within meters for which traditional accelerators require several kilometers, and require precise control of the current profiles of accelerated bunches \cite{ref_FACET,ref_FACET_2}.

In this work we demonstrate a general model-independent feedback algorithm which can be used for online multi-objective optimization for a wide range of particle accelerator problems by running multiple competing feedback loops simultaneously at multiple time scales. We demonstrate the strength of our algorithm by solving the multi-objective optimization problem of minimizing transverse electron beam size ($\sigma_x,\sigma_y$) at the end of the electron beam line of CERN's Advanced Proton Driven Plasma Wakefield Acceleration Experiment (AWAKE) while maintaining a design orbit. In this approach, minimizing beam size was equivalent to minimizing emittance because only
the optics at the beginning of the beam line were adjusted. Control of the electron bunch at the end of this beamline is important for the PWFA experiments that take place, sending electron bunches into the wakes generated by 400 GeV protons from CERN's Super Proton Synchrotron (SPS) accelerator \cite{ref_AWAKE_e_ACCEL}, labeled as the electron-proton overlap region in figure \ref{fig:AWAKE_e}.

The first step was to demonstrate an ability to control the trajectory of the electron beam in the x-plane through 10 beam position monitors (BPM) starting with the BPMs labeled as BPM 039 through BPM 351 in figure \ref{fig:AWAKE_e}, which will be referred to as BPM number 1 - 10 below. Control of the trajectory was achieved by an iterative tuning of 10 horizontal steering magnets located directly in front of each BPM. 

The iterative algorithm is based on a recently developed form \cite{ref_ES_2} of a model-independent feedback tuning algorithm designed for high dimensional noisy systems \cite{ref_ES_1}. Given a noisy measurement $\hat{C}$ of an analytically unknown cost function to be minimized, $C$, by adjusting a group of coupled parameters $\mathbf{p} = (p_1,\dots,p_m)$, the method proceeds iteratively according to
\begin{equation}
    p_i(n+1) = p_i(n) + \Delta_t \sqrt{\alpha \omega_i}\cos \left (\omega_i n \Delta_t + k \hat{C}(n) \right ), \label{ES_step}
\end{equation}
where each parameter $p_i$ has a unique dithering frequency $\omega_i$, $\alpha$ controls the dither size and may be increased to escape local minima, $k$ is a feedback gain, increasing which may speed up convergence, and $\Delta_t \ll 1$ is chosen small enough such that (\ref{ES_step}) is a finite difference approximation of
\begin{equation}
    \frac{dp_i(t)}{dt} = \sqrt{\alpha \omega_i}\cos \left (\omega_i t + k \hat{C}(\mathbf{p},t) \right ),
\end{equation}
which results in minimization of the unknown function $C$ (although only having access to its noise-corrupted measurements), according to the average dynamics \cite{ref_ES_2,ref_ES_1}
\begin{equation}
    \frac{d\bar{\mathbf{p}}}{dt} = -\frac{k\alpha}{2}\nabla_{\mathbf{p}} C(\mathbf{p},t).
\end{equation}

In this single objective case the goal was to minimize the X root mean square error (RMSE) between the trajectory and a target trajectory, as given during each iteration, $n$, by
\begin{equation}
    X_\mathrm{RMSE}(n) = \sqrt{\frac{1}{10}\sum_{i=1}^{10}\left (\mathrm{BPM}_i(n) - \mathrm{BPM}_{i,o} \right )^2},
\end{equation}
where $\mathrm{BPM}_i(n)$ was the BPM measurement recorded at step $n$ and $\mathrm{BPM}_{i,o}$ was the desired orbit BPM reading. The algorithm proceeded by first introducing offsets in all steering magnets, $M_i(1)$, in order to create a large deviation from the target orbit and recording $X_\mathrm{RMSE}(1)$. Each magnet was then adjusted iteratively according to (\ref{ES_step}) and a new value, $X_\mathrm{RMSE}(2)$, was recorded. The tuning parameters used were: $\alpha = 0.15$, $k=0.2$, the $\omega_i$ values were evenly distributed between 100 and 175 so that all of the frequencies were distinct and no single frequency was an integer multiple of the other, and $\Delta_t = 2\pi / (10 \times \max \{\omega_i \})$. The results of this optimization are shown in figure \ref{fig:traj_e}. It is clear that despite a large energy and therefore trajectory jitter due to the beam line's Klystron, the algorithm was able to achieve convergence within 30 steps.

The next objective was to utilize two such feedbacks simultaneously, at two different time scales, for multi-objective optimization. One feedback was used to adjust two solenoids and three quadrupole magnets directly following the electron beam source, to minimize electron beam size at the end of the beam line. For this feedback we recorded beam images, as shown in figure \ref{fig:emittance_e}, projected them onto the x and y axes, and then fit Gaussian distributions of the form
\begin{equation}
	f_x(x) = A_x e^{ -(x-\mu_x)^2/2\sigma_x^2 }, \quad f_y(y) = A_y e^{ -(y-\mu_y)^2/2\sigma_y^2 },
\end{equation}
and estimated beam size based on $\sigma_x$ and $\sigma_y$  as
\begin{equation}
    \sigma_{xy} = \sqrt{\sigma_x^2 + \sigma_y^2},
\end{equation}
which was the cost to be minimized according to (\ref{ES_step}). The tuning parameters used for this feedback were: $\alpha = 0.1$ for each Solenoid and $\alpha=0.05$ for each quadrupole magnet, $k=1.0$, and $\omega_i$ evenly distributed between 100 and 175, and $\Delta_{t,2} = 2\pi / (31 \times \max \{\omega_i \})$. These adjustments caused trajectory changes, driving the beam off of the beam-size detector or close to the available aperture. 

A second feedback, which has already been described above, running on a faster time scale, was simultaneously run to continuously adjust 10 steering magnets, based on 10 beam position monitor readings, to maintain a prescribed reference trajectory for the beam. The overall setup is shown in figure \ref{fig:AWAKE_e}. The beam size feedback had a value of $\Delta_{t,2} < \Delta_t/3$, and therefore ran at a different time scale from the trajectory maintaining feedback, 3 times slower, allowing for the steering magnets to quickly compensate for trajectory deviations caused by changes in the solenoid and quadrupole magnets.

The simultaneous evolution of all of the 15 components is shown in figure \ref{fig:emittance_params_e} alongside the evolution of the two objectives. By running the two feedback loops simultaneously we were able to solve the multi-objective problem:
\begin{equation}
    C(n) = w_1 X_\mathrm{RMSE}(n) + w_2 \sigma_{xy}(n),
\end{equation}
with weights $w_1 = 0.2$, $w_2 = 1.0$. This approach was able to continuously re-adjust steering magnets attempting to maintain the target beam orbit, thereby keeping the beam on the screen and allowing the second feedback to minimize the beam size, resulting with a decrease over 2$\times$ more than what had previously been achieved, as shown in figure \ref{fig:emittance_e}.

The experiment also demonstrated possible limitations of this model-independent feedback approach. Because the cost function was only based on maintaining a trajectory and minimizing beam size, it did not penalize beam loss. In figure \ref{fig:emit_size} we plot the evolution of the beam size, $\sigma_{xy}$ together with the evolution of detector image sums, as well as a measurement of intensity given by sum$/\sigma_{xy}$. As beam size is decreased and intensity is increased, it appears that we are also losing some of the beam. Future work would take into account additional information about the beam, including beam losses.

This demonstration of the use of multiple feedbacks at various time scales for simultaneous multi-objective optimization for particle accelerators is a very general approach that is robust to noise and time-variation of components and beam properties. This general approach can be useful for a wide range of accelerator problems. One important example of this may be maintaining a desired beam trajectory and maximizing FEL output power while adjusting bunch energy, length, and charge. Such adjustments are frequently made at FELs when switching between different users, all of which have unique X-ray frequency and pulse duration requirements, sometimes require many hours of expert tuning for large changes, and must be continuously adjusted during steady state operation due to beam and accelerator component fluctuations. The same is true of precisely tuned bunch current profiles at PWFAs.

\nocite{*}
\bibliography{AWAKE_Multi_ES}% Produces the bibliography via BibTeX.

\end{document}